\documentclass[twocolumn]{aa}
\usepackage{graphicx}
\usepackage{txfonts}
\begin{document}
   \title{On the structure of globular cluster systems in elliptical galaxies}


   \author{Kenji Bekki
          \inst{1}
          \and
          Duncan  A. Forbes\inst{2}
          }

   \offprints{K. Bekki}

   \institute{School of Physics, University of New South Wales,
              Sydney 2052, NSW, Australia\\
              \email{bekki@bat.phys.unsw.edu.au}
         \and
             Centre for Astrophysics and Supercomputing,
              Swinburne University of Technology, Mail H39, PO Box 218,
              Hawthorn, VIC 3122, Australia \\
             \email{dforbes@astro.swin.edu.au}
             }

   \date{}

   \abstract{
It has long been known that the radial density profiles of
globular cluster systems (GCSs) in elliptical galaxies
vary with the total luminosities of their host galaxies.
In order to elucidate the origin of this structural non-homology
in GCSs, we numerically investigate the structural properties
of GCSs in elliptical galaxies formed from a sequence of 
major dissipationless galaxy merging.  
We find that the radial density profiles of GCSs 
in elliptical galaxies  become progressively flatter as the galaxies
experience more major merger events.
The density profiles of GCSs in ellipticals are well described as power-laws
with slopes (${\alpha}_{\rm gc}$) ranging from $-2.0$ to $-1.0$.
They are flatter than,
and linearly proportional to, the slopes (${\alpha}_{\rm s}$)
of the stellar density profiles
of their host galaxies.
We also find that the GCS core radii ($r_{\rm c}$) of the density profiles 
are larger in ellipticals that experienced more mergers.
By applying a reasonable scaling relation between luminosities and sizes of galaxies
to the simulation results,
we show that ${\alpha}_{\rm gc} \approx -0.36 M_{\rm V}-9.2$, 
$r_{\rm c} \approx -1.85 M_{\rm V}$,
and ${\alpha}_{\rm gc} \approx 0.93 {\alpha}_{\rm s}$,
where $M_{\rm V}$ is the total $V$-band absolute magnitude of a galaxy.
We compare these predictions with observations
and discuss their physical meaning. 
We suggest that the origin of structural non-homology of GCSs 
in ellipticals can be understood in terms of the  growth
of ellipticals via major dissipationless galaxy merging.
   }

   \maketitle
%

\section{Introduction}
The structural properties of globular cluster systems (GCSs)
in galaxies have long been suggested  to 
contain fossil information on the early dynamical histories
of galaxies and has accordingly been investigated both
observationally and theoretically (e.g., Harris 1986, 1991).
In particular,  radial density profiles of GCSs  and their
correlations with physical properties of their host galaxies
have been discussed
in terms of formation and evolution of elliptical galaxies 
(e.g., Harris 1986; Zepf \& Ashman 1993; Ashman \& Zepf 1998;
Forbes et al. 1996, 1997; 
Bekki et al. 2002; 2003). 
In particular, observational studies found that the 
GCSs in early-type galaxies were less centrally concentrated than
the stellar light of the host galaxy 
(e.g., Harris \& Racine 1979;  Strom et al. 1981; Grillmair et
al. 1994; Forbes et al. 1996), and that the outer slope of the GCS
density profile correlated with the host galaxy luminosity 
(e.g., Ashman \& Zepf 1998; Harris 1986).

Major galaxy merging between equal-mass disk galaxies 
has long been considered to be a promising formation scenario 
for elliptical galaxy formation (e.g., Toomre 1977).
Although this galaxy formation scenario has been discussed in many
contexts (e.g., the color-magnitude relation),
it has not been extensively
investigated in terms of whether it can self-consistently explain
the observed properties of GCSs around ellipticals.
So far, only the observed  higher specific frequency ($S_{\rm N}$) and
bimodal color distributions of GCSs have been discussed in terms
of hierarchical galaxy formation   (Beasley et al. 2002). 
It is thus unknown what implications the observed {\it structural
properties of GCSs} in ellipticals has for galaxy formation.

The purpose of this paper is to propose
that the radial density profiles  of GCSs in elliptical galaxies
have fossil information on how often ellipticals 
have experienced major merger events in their dynamical histories.
By using dissipationless N-body
simulations of major merging of galaxies,
we investigate how the structural properties of GCSs in elliptical
galaxies change as they grow by successive (gas-free) mergers.
We compare our theoretical predictions with 
observations to discuss the  origin of the observational
trends. We also briefly discuss the role of GC accretion. 
In particular, we focus on the luminosity-dependence of
the GCS slope (${\alpha}_{\rm gc}$) 
which implies a ``non-homology'' in the structural properties
of GCSs. 

   \begin{figure}
   \centering
   \includegraphics[width=8.5cm]{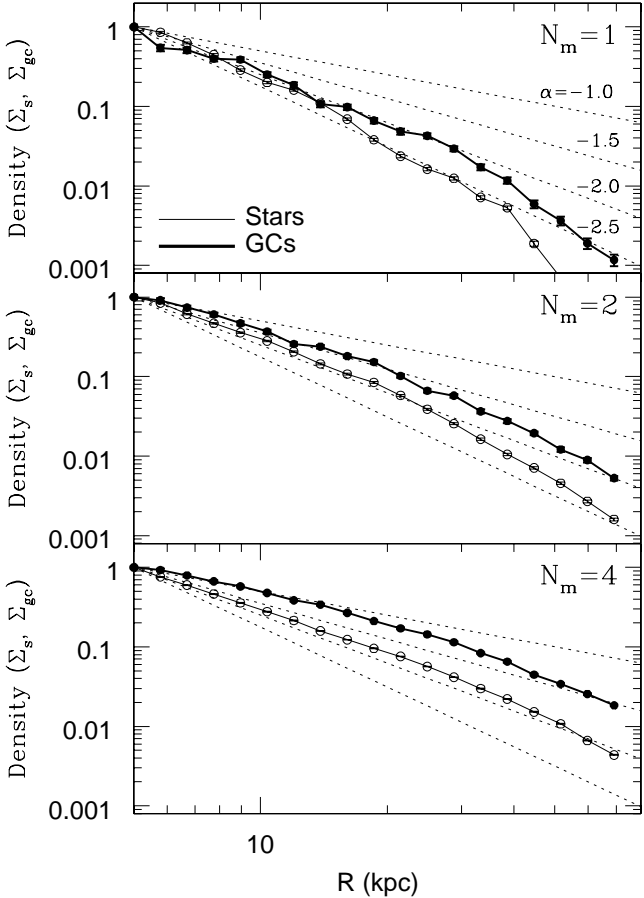}
      \caption{
Dependences of projected number distributions of stars (thin)
and GCs (thick) in merger remnants (i.e., elliptical galaxies)
on  the total number of major
merger events ($N_{\rm m}$) which an elliptical experienced during its formation.
For clarity, the density distributions are normalized to
their central values.
Thin dotted lines represent power-law slopes ($\alpha$)
of $\alpha$ = $-2.5$, $-2.0$, $-1.5$, and $-1.0$.
Note that the density profiles of GCSs become flatter
for larger $N_{\rm m}$, i.e. more mergers.
              }
         \label{FigVibStab}
   \end{figure}

\section{Model}

We  investigate the dynamical evolution of GCSs 
of elliptical galaxies formed from sequential dissipationless major merging
based on numerical simulations  
carried out on the GRAPE
board (Sugimoto et al. 1990) in GRAPE 5 systems.
We here stress that the roles of minor merging/accretion of GC in
shaping radial density profiles of GCSs in galaxies are not 
extensively investigated in the present models.
We adopt the following  merging scheme to model
elliptical galaxy formation through sequential  major equal-mass disk mergers.  
First an elliptical galaxy is formed by major merging
between two equal-mass disk galaxies.  
We refer to the remnant elliptical galaxy as the 1st-generation 
elliptical.
This 1st-generation elliptical then merges with other 
1st-generation one to form the 2nd-generation
elliptical.
This sequential major merging can repeat for $N_{\rm max}$ times, and the structural 
properties of stellar components and GCSs are investigated
for each  {\it i}th generation elliptical ($i \le N_{\rm max}$).
The total particle number used 
is 27670 for  the simulations of the 1st-generation  elliptical
and 221360 for those of the 4th-generation one (i.e.,  $N_{\rm max}$ of 4). 
The initial total particle number for GCs is set to be 1000,
because an order of 1000 (not 10 or 100) is necessary to make robust
predictions for the radial profiles of structure and kinematics of GCSs 
in galaxies (Bekki et al. 2005).

Since the numerical methods and techniques we employ for modeling
dynamical evolution of galaxy mergers with GCs 
have already been described in detail 
elsewhere (Bekki et al. 2002, 2005), we give only  a brief review here. 
The progenitor disk galaxies that take part in a merger 
(for the 1st-generation elliptical) are assumed to 
have a dark halo, a bulge, a stellar halo,  a thin exponential disk,
and a GCS.
The total disk mass and size are $M_{\rm d}$ and $R_{\rm d}$, respectively. 
Henceforth, all masses are measured in units of
$M_{\rm d}$ and  distances in units of $R_{\rm d}$, unless otherwise specified. 
We adopt the density distribution of the NFW
halo (Navarro, Frenk \& White 1996) suggested from CDM simulations:
 \begin{equation}
 {\rho}(r)=\frac{\rho_{0}}{(r/r_{\rm s})(1+r/r_{\rm s})^2},
 \end{equation} 
where  $r$, $\rho_{0}$, and $r_{\rm s}$ are
the spherical radius,  the central density of a dark halo,  and the scale
length of the halo, respectively.  
The dark matter to disk mass ratio is fixed at 9 for all models.
The value of $r_{\rm s}$ (typically $\sim$ 3$R_{\rm d}$) is chosen such that
the rotation curve of the disk is reasonably consistent with
observations for a given bulge mass.
We adopt
the $R^{1/4}$  profile for the bulge with the 
mass of 0.17 and a scale length of 0.04$R_{\rm d}$.
The radial ($R$) and vertical ($Z$) density profiles 
of the  disk are  assumed to be
proportional to $\exp (-R/R_{0}) $ with scale length $R_{0}$ = 0.2$R_{\rm d}$
and to  ${\rm sech}^2 (Z/Z_{0})$ with scale length $Z_{0}$ = 0.04$R_{\rm d}$
in our units, respectively.

The GCSs in the progenitor spirals have a spherical distribution with
a density profile of ${\rho}(r)$ $\propto$ $r^{ {\alpha}_{\rm gc,i} }$. 
We adopt ${\alpha}_{\rm gc,i} = -3.5$ for most models (however 
the dependence of the final radial profiles of GCSs in mergers
on ${\alpha}_{\rm gc,i}$ is also investigated).
The adopted  density profile of ${\rho}(r)$ $\propto$ $r^{-3.5}$ is
consistent with that observed for the Galactic GCS 
(Djorgovski \&  Meylan 1994). The GCS of M31 reveals a similar
surface density distribution to the Milky Way GCS
(e.g. Battistini et al. 1993). Very little is known about the
density profile of GCSs in other spirals. 
The velocity dispersion of a GCS is assumed to be isotropic
which is consistent with the observations of the Galactic GCS
(e.g., Freeman 1993).

The simulations
are dissipationless so no new GCs form in the merging
process. Thus our simulations are more appropriate for relatively
gas-free mergers which may be expected to occur at some late epochs. 
Recently, observational studies by Bell et al. (2005)
have demonstrated that 
present day spheroidal galaxies with $M_{\rm V} < -20.5$ 
on average have undergone
between 0.5 and 1 major ``dry merger'' (i.e., gas-poor mergers)
since z $\sim$ 0.7.
This observational paper thus  suggests that the dissipationless
merger models adopted in the present paper can be  reasonable
for elliptical galaxy formation.
Proto-GCs have been observed to form in late epoch
mergers (e.g. Whitmore \& Schweizer 1995),
which implies that some fraction
of GCs in ellipticals could be quite young.
However, latest observational studies
(e.g. Cohen et al. 1998, 2003;
Forbes et al. 2001; Kuntschner et al. 2002; Beasley et al. 2004;
Strader et al. 2004; Pierce et al. 2005), in which
more reliable age estimation of GCs can be done based on
the higher resolution spectra and the improved stellar population
synthesis model, 
have demonstrated that the vast bulk of GCs
in early-type galaxies are very old, i.e. $\ge$ 10 Gyrs.
These suggest that 
pre-existing GCs dominate in old merger remnants.

In all of the simulations, the orbit of the two galaxies is set to be
initially in the $xy$ plane and the distance between
the center of mass of the two 
is  assumed  to be 6 in our units. 
The pericenter distance ($r_{\rm p}$) and the eccentricity ($e_{\rm p}$)
in a  merger
are assumed  to be free parameters that control
orbital energy and angular momentum of the merger. 
The spin of each galaxy in a merger
is specified by two angles $\theta_{i}$ and
$\phi_{i}$, where suffix  $i$ is used to identify each galaxy.
Here $\theta_{i}$ is the angle between the $z$ axis and the vector of
the angular momentum of a galaxy, and 
$\phi_{i}$ is the azimuthal angle measured from the $x$ axis to
the projection of the angular momentum vector of a galaxy onto the $xy$ plane. 
We specifically show four
different and representative models  of merger sequences with $N_{\rm max}=4$ 
in which merger pairs take one of the following orbital configurations
(i.e., galaxy inclinations with respect to the orbital plane):
A prograde-prograde model represented by ``PP''
with $\theta_{1}$ = 0, $\theta_{2}$ = 30, $\phi_{1}$ = 0,
$\phi_{2}$ = 0, 
$r_{\rm p}$ = 1.0, and  $e_{\rm p}$ = 0.72 
a retrograde-retrograde (``RR'') 
with $\theta_{1}$ = 180,
$\theta_{2}$ = 150, $\phi_{1}$ = 0, $\phi_{2}$ = 0, 
$r_{\rm p}$ = 1.0, and  $e_{\rm p}$ = 0.72 
and a highly
inclined model (``HI'') with 
$\theta_{1}$ = 30, $\theta_{2}$ =
120, $\phi_{1}$ = 90, and $\phi_{2}$ = 180,
$r_{\rm p}$ = 1.0, $e_{\rm p}$ = 1.0
and the low orbital angular momentum model (``LA'')
with 
$\theta_{1}$ = 30, $\theta_{2}$ =
120, $\phi_{1}$ = 90, and $\phi_{2}$ = 180,
$r_{\rm p}$ = 0.2, $e_{\rm p}$ = 1.0.

In analyzing the projected  radial density profiles of GCs (${\Sigma}_{\rm gc}$)
and stars (${\Sigma}_{\rm s}$), we assume that
they are approximated as ${\Sigma}_{\rm gc} \propto R^{{\alpha}_{\rm gc}}$
and ${\Sigma}_{\rm s} \propto R^{{\alpha}_{\rm s}}$,
where $R$ is the projected distance from the center of a galaxy.
We derive the slopes of ${\alpha}_{\rm gc}$ and ${\alpha}_{\rm s}$ 
and the difference of the two 
${\alpha}_{\rm gc}-{\alpha}_{\rm s}$,
for $5\le R {\rm (kpc)} \le 20$ so that the derived values
can be directly compared with observations.
We determine a core radius ($r_{\rm c}$) for the GCS 
by adopting an isothermal profile of the form 
${\Sigma}_{\rm gc} \propto {(r^2+{r_{\rm c}}^2)}^{-1}$, which can
be directly compared to the observations  of Forbes et al. (1996).
For an isothermal profile $r_{\rm c}$ $\approx $ $0.22R_{\rm h,gc}$,
where $R_{\rm h,gc}$ is a half-number radius of the GCS.
In order to estimate the total $V$-band magnitude of a merger remnant,
we assume an initial disk mass of $M_{\rm d}=4.0 \times 10^{10} M_{\odot}$,
an initial disk size of $R_{\rm d}=14$ kpc and 
$M_{\rm d}/L_{\rm V}=5.0$, where $L_{\rm V}$ is the total $V$-band
luminosity of the initial disk. 
The remnant's total mass is  $2^{N_{\rm m}} \times M_{\rm d}$,
where $N_{\rm m}$ is total number of major merger events that
the remnant elliptical experienced.

All the calculations related to the above dissipationless evolution
have been carried out on a GRAPE
board (Sugimoto et al. 1990) in the GRAPE 5 system
at the National Astronomical Observatory of Japan. 
The parameter of gravitational softening for
stellar particles is set to be fixed at 0.039 in our
units (0.68 kpc). The time integration of
the equation of motion is performed by using the 2nd-order leap-flog method.

   \begin{figure}
   \centering
   \includegraphics[width=8.5cm]{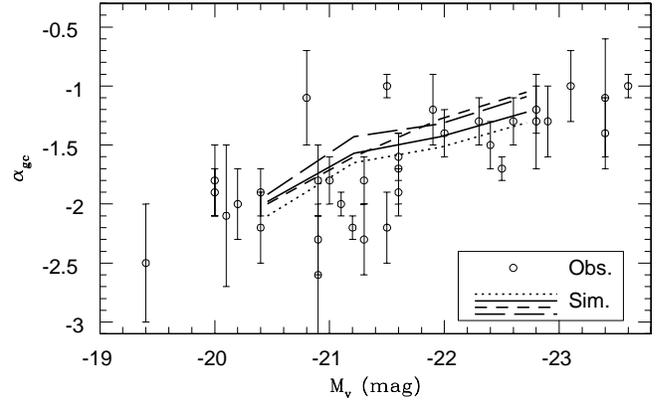}
      \caption{
Comparison between the observed ${\alpha}_{\rm gc}-M_{\rm V}$
relation and the simulated one.
The observational data, represented by open circles, come from
the table in the Appendix of Ashman \& Zepf (1998).
Solid, dotted, short-dashed, and long-dashed lines
show the results of the HI, PP, RR, and LA models, respectively.
Note that both simulations and observations show a tendency
for flatter density profiles in  brighter ellipticals.
              }
         \label{FigVibStab}
   \end{figure}

   \begin{figure}
   \centering
   \includegraphics[width=8.5cm]{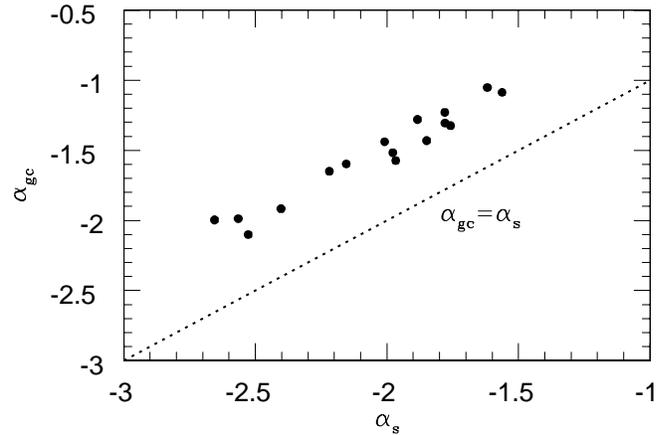}
      \caption{
The dependence of the GCS profile slope (${\alpha}_{\rm gc}$) and
that for the galaxy stars (${\alpha}_{\rm s}$)
in 16 simulated merger remnants, i.e. elliptical galaxies.
The dotted line indicates 
${\alpha}_{\rm gc} = {\alpha}_{\rm s}$.
              }
         \label{FigVibStab}
   \end{figure}

   \begin{figure}
   \centering
   \includegraphics[width=8.5cm]{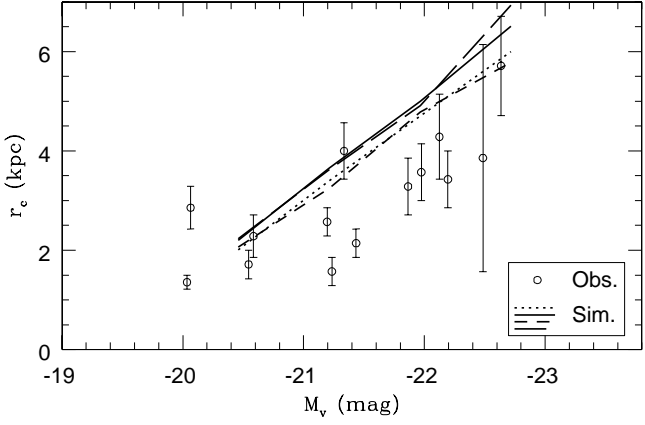}
      \caption{
Comparison between the observed
$r_{\rm c}-M_{\rm V}$
and the simulated ones.
Here $r_{\rm c}$ is the core radius of
a GCS's distribution.
The observational data represented by open circles come from
Forbes et al. (1996).
Solid, dotted, short-dashed, and long-dashed line
show the results of the HI, PP, RR, and LA models, respectively.
              }
         \label{FigVibStab}
   \end{figure}

\section{Results}
\subsection{Structural non-homology}

Figure 1 plots the radial density profiles of stars and GCs
in ellipticals variation with $N_{\rm m}$ in the HI model sequence.
The reference GC profile with ${\alpha}_{\rm gc}=-2.5$ in this figure
corresponds to the Galactic GC profile (Zinn 1985) which
shows ${\alpha}_{\rm gc} \sim -2$ out to $R=10$ kpc
and ${\alpha}_{\rm gc} \sim -3$ beyond $R=20$ kpc (Harris 1976).
This figure shows that up to $N_{\rm m}$ = 4, the ellipticals have 
positive ${\alpha}_{\rm gc}-{\alpha}_{\rm s}$ (e.g. 0.58 for $N_{\rm m}=1$),
and thus show flatter profiles for GCs than for field stars, which is consistent
qualitatively with observations (e.g., Harris 1986, 1991). 
The reason for this is that GCs in a merger  are more `puffed up'
relative to the stars after they inevitably absorb orbital angular momentum and kinetic
energy during the merger,  
because most GCs are initially located in the galactic outer parts
where such absorption 
is most likely to occur.
The radial density profile of a GCS in an elliptical
is always flatter in the inner region ($R<20$ kpc)
than in the outer one ($R\ge20$ kpc). Again, qualitatively
consistent with observations (e.g., Forbes et al. 1996).

Both stellar and GC profiles in  an elliptical become progressively flatter 
as  the  elliptical experiences more major merger events 
(i.e., larger $N_{\rm m}$).
For example, in the HI model 
${\alpha}_{\rm gc}$ is estimated as $-1.99$ for $N_{\rm m}=1$
and $-1.23$ for $N_{\rm m}=4$, which implies
that {\it the slope ${\alpha}_{\rm gc}$ of a GCS in
an elliptical  has  fossil information
on how many times the elliptical has experienced dissipationless major merger
events in its formation history.} 
The derived slopes for 
ellipticals are generally flatter than that observed 
for the Galactic GCS which has ${\alpha}_{\rm gc} \approx -2.5$ (Zinn 1985). 
This suggests that major merging can transform the initially steeper
density profiles of GCSs in spirals into the flatter ones observed in
ellipticals. The general trends shown in Figure 1 do not depend on model parameters.

Figure 2 shows how ${\alpha}_{\rm gc}$ 
depends on the total $V$-band magnitude of ellipticals 
for four different sets of models (i.e., PP, RR, HI, and LA).
Irrespective of the models,
GCSs in brighter ellipticals show flatter
density profiles (i.e., larger ${\alpha}_{\rm gc}$), which is
reasonably consistent  with observations.
This is due to the fact that ${\alpha}_{\rm gc}$
becomes larger each time a major merger occurs.
A least square fit to the simulation data shown in
Figure 2 gives 
${\alpha}_{\rm gc} \approx -0.36M_{\rm V}-9.2$,
which is similar to the observed relation of 
${\alpha}_{\rm gc} \propto -0.3 M_{\rm V}$ (Harris 1986).
This result suggests that the origin of the luminosity-dependent
${\alpha}_{\rm gc}$ (Harris 1991; Ashman \& Zepf 1998) can be 
understood in terms of the growth of ellipticals
via major dissipationless mergers.
It should be however noted, that the simulated range of
${\alpha}_{\rm gc}$, 
for a given luminosity, is quite  narrow so that 
the present models can not simply explain the scatter 
observed for ${\alpha}_{\rm gc}$ 
(in particular, for $-22<M_{\rm V}<-21$). 

Figure 3 shows a positive correlation between 
${\alpha}_{\rm gc}$ and ${\alpha}_{\rm s}$
(${\alpha}_{\rm gc} \approx 0.93{\alpha}_{\rm s}+0.37$), which means that
ellipticals with flatter distributions of stars are highly
likely to show flatter distributions of GCSs.
Figure 3 also shows that ${\alpha}_{\rm s}$ is always smaller
than ${\alpha}_{\rm gc}$ (i.e., ${\alpha}_{\rm gc}-{\alpha}_{\rm
s} \sim 0.4$) and thus confirms
that elliptical galaxies formed by major merging can have GCSs with
the density profiles flatter than those of stars. 
The density profiles of the GCSs can be fit by a 
power-law with slope ${\alpha}_{\rm gc} \approx -0.4$ for 
the inner region of $0.5 \le R \le 10$ kpc.
However the shallower profiles of GCSs
in the inner regions can not be 
regarded as ``flat cores'' (${\alpha}_{\rm gc} \sim 0$) which
appear to be common  
in  the GCSs of giant ellipticals (e.g., Harris 1986, 1991; Forbes et
al. 1996). 

Figure 4 shows a clear trend of larger GCS core radii ($r_{\rm c}$) 
for brighter ellipticals, which is consistent qualitatively with
the observational results of Forbes et al. (1996) for the overall
GCS. 
The derived trend is due to the fact that $r_{\rm c}$ becomes larger
by a factor of $\sim$ 1.4 after each 
major merger.  
We can estimate the luminosity-dependence
of $r_{\rm c}$ as $r_{\rm c} \propto  -1.41M_{\rm V}$
for the observational data with $M_{\rm V}\le-20.5$ of 
Forbes et al. (1996), 
and as $r_{\rm c} \propto  -1.85M_{\rm V}$
for the simulation data set shown in Figure 4. 
The derived luminosity-dependence of $r_{\rm c}$ is therefore 
steeper than the observed one. 

   \begin{figure}
   \centering
   \includegraphics[width=8.5cm]{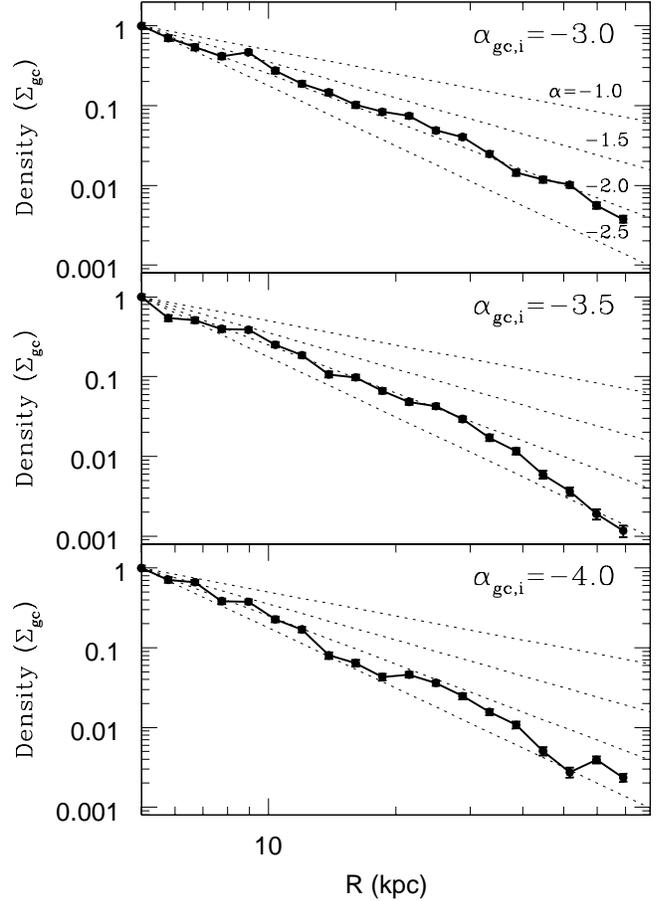}
      \caption{
Dependences of projected number distributions of 
GCs in merger remnants (i.e., elliptical galaxies)
on the initial slope of radial density profiles of GCSs in
spirals (i.e., ${\alpha}_{\rm gc, i}$) for the models
with $N_{\rm m}=1$.
For clarity, the density distributions are normalized to
their central values.
Thin dotted lines represent power-law slopes ($\alpha$)
of $\alpha$ = $-2.5$, $-2.0$, $-1.5$, and $-1.0$.
Note that there are no remarkable differences between
the three models.
              }
         \label{FigVibStab}
   \end{figure}

   \begin{figure}
   \centering
   \includegraphics[width=8.5cm]{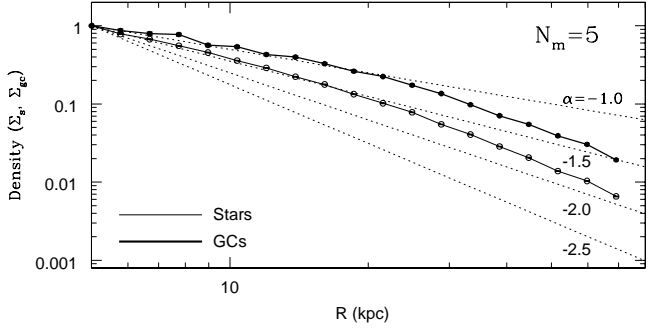}
      \caption{
The projected number distributions of stars (thin)
and GCs (thick) in merger remnant (i.e., elliptical galaxies)
with $N_{\rm m}=5$.
For clarity, the density distributions are normalized to
their central values.
Thin dotted lines represent power-law slopes ($\alpha$)
of $\alpha$ = $-2.5$, $-2.0$, $-1.5$, and $-1.0$.
              }
         \label{FigVibStab}
   \end{figure}

\subsection{Robustness of the results}

\subsubsection{Dependence on ${\alpha}_{\rm gc,i}$}

Although we adopt the slope of the Galactic GCS as the initial
radial profile for GCSs in our progenitor spirals (i.e., ${\alpha}_{\rm gc,i}=-3.5$),
our results do not depend strongly on the initial profile slope.
Figure 5 shows how the radial profiles of GCSs depend on
${\alpha}_{\rm gc,i}$ for the models with $N_{\rm m}=1$.
It is clear from this figure that there is no remarkable
difference in the profiles between the three models
for $R<20$ kpc: The difference of the final ${\alpha}_{\rm gc}$ 
between the models with ${\alpha}_{\rm gc,i}=-3.0$ and $-4.0$
becomes at most  $\sim 0.5$, if we fit the profiles to
the power-law ones for $R<20$ kpc.
It should be also stressed that the model with 
${\alpha}_{\rm gc,i}=-3.5$ and $-4.0$
shows a slightly steeper profile for  the outer parts of the
merger remnants ($R>40$ kpc) compared with the model
with ${\alpha}_{\rm gc,i}=-3.0$.
This tendency can be seen in other models with different
orbital configurations.

The derived weak dependence on ${\alpha}_{\rm gc,i}$ is due
to the fact that violent relaxation during major galaxy merging
can effectively wipe
out the original density distributions of GCSs for
$-4.0 \le {\alpha}_{\rm gc,i} \le -3.0$.
The resulting profiles of ${\alpha}_{\rm gc} \sim -2$ of GCSs
in merger remnants with $N_{\rm m}=1$
follow their dark matter halos, and furthermore 
do not depend on initial orbital configurations of galaxy merging.
These results, combined with those shown in Figure 1,
imply that the present results on the dependence of
${\alpha}_{\rm gc}$ on $N_{\rm m}$ (thus on $M_{\rm V}$)
does not depend strongly on ${\alpha}_{\rm gc,i}$  and thus can be regarded
as fairly robust.

\subsubsection{${N}_{\rm m} > 4$}

It is of interest to investigate whether the derived dependence
of ${\alpha}_{\rm gc,i}$ on $N_{\rm m}$ can be seen in models
with $N_{\rm max} > 4$ and thereby confirm that results are true
for a possible range of $N_{\rm m}$.
We note that the most massive ellipticals, with $M_{\rm V} < -23$
mag, are typically cD galaxies at the centres of clusters, for which accretion of GCs from
cluster member galaxies are important (e.g., Bekki et al. 2003)
and may well play a role in determining the radial profiles of
GCSs. The luminosity range (for $N_{\rm max} \le 4$) 
in the present simulations is thus appropriate for most
ellipticals but not cD galaxies.

Figure 6 shows the radial density profiles of stars and GCs in
the model with $N_{\rm m}=5$  for the PR orbital configuration.
It is clear from this figure and Fig. 1 that the profile
of GCS becomes even flatter than that in the model with $N_{\rm
m} \le 4$.
This result can be seen in other models with different orbital configurations.
For example, ${\alpha}_{\rm gc}$ is estimated as $-0.77$ for
the PR and $-1.03$ for the PP models.
These results confirm that radial density profiles of GCSs of ellipticals
formed by sequential major merging depend strongly on $N_{\rm m}$. 
We note that for $N_{\rm max} \ge 5$ and  
$M_{\rm d}=4.0 \times 10^{10} M_{\odot}$, the resulting merger remnant has  
$M_{\rm V} < -24$ mag and thus few if any observational counterparts.

\subsubsection{Spatial coverage of GCSs}

It should be noted that the spatial coverage and number of GCs
observed for the estimation of ${\alpha}_{\rm gc}$ (e.g., Ashman \& Zepf 1998)
varies from galaxy to galaxy.
The spatial coverage
of the simulation data, on the other hand, is fixed at $5\le R \le 20$kpc
for the ${\alpha}_{\rm gc}$ estimation,
which suggests that the present compassion is not fully self-consistent.
It is accordingly important for the present study to show 
how the ${\alpha}_{\rm gc}$ values depend on the spatial coverage.
It is found that ${\alpha}_{\rm gc}$ is $\approx -2.0$ for
$5\le R \le 20$kpc, 
$\approx -1.6$ for $5\le R \le 10$kpc,
and $\approx -2.4$ for $10\le R \le 20$kpc in the HI model with
$N_{\rm m}=1$.
This result suggests  that  (1) ${\alpha}_{\rm gc}$
can be larger (i.e., GCS density profiles are flatter)
if ${\alpha}_{\rm gc}$ is derived for the inner
regions of ellipticals and (2) ${\alpha}_{\rm gc}$ can
be smaller (i.e., GCS density profiles are steeper)
if ${\alpha}_{\rm gc}$ is derived for the outer parts of
ellipticals. Given the fact that these tendencies can be
seen in other models (PP, RR, and LA) the present results imply that
a more careful comparison between observations and
simulations is necessary in deriving physical meaning
of the  GCS density profiles in ellipticals.

\section{Discussion}


In the present study we suggest that sequential major merging
has the physical effect of flattening the radial density profiles of GCSs
so that the radial profiles of GCSs in ellipticals 
contain fossil information about 
their past merger history. 
Neither of the above two points has been previously suggested 
by theoretical studies. 

The GCSs in ellipticals are thought to be subject to dynamical
destruction processes (e.g. Baumgardt 1998; Fall \& Zhang 2001;
Vesperini et al. 2003), which are not included in this work.  
However there is currently little convincing observational evidence  
for destruction of GCs in ellipticals, e.g. Harris et al. (1998)
found no variation of the bright end of the GCLF with
galactocentric radius in M87. 
Vesperini et al. (2003)
also found no variation in the GCLF with radius in M87, which 
they claimed was consistent with dynamical destruction if
the GC mass function initially had a bell-shaped distribution. 

To date, theoretical studies which focus on GC dynamical destruction
have not yet investigated the luminosity dependence
of the GCS density slope (${\alpha}_{\rm gc}$). 
Thus it remains to be seen whether GC destruction can account for
this dependence, which is modelled here by sequential merging.

Destruction processes are expected to be
stronger in the galaxy inner regions and thus 
play some role in determining the inner core radius of the
GCS. However it is difficult to understand how such processes could
explain the trend of r$_c$ with galaxy luminosity given that destruction
processes are expected to be {\it less} efficient in the central regions
of more {\it massive} ellipticals (Murali \& Weinberg 1997).

Although the present sequential merger model can qualitatively
explain some of the observed GCS structural properties, it 
has the following three
disadvantages:
Firstly, it remains unclear why
GCSs in  some intermediate
luminosity ellipticals ($M_{\rm v}<-21.5$)
can show steeper (${\alpha}_{\rm gc} < -2.0$) profiles,
which are not fully consistent with the observation
(See Fig. 2). 
Secondly, the simulated core radius ($r_{\rm c}$) of a GCS
for a given luminosity is appreciably larger than the observed
one (Note that the derived luminosity-dependence of $r_{\rm c}$ is
steeper than the observed one in Figure 4). Thirdly, although the derived trend that
${\alpha}_{\rm gc}$ is always larger than ${\alpha}_{\rm s}$
is broadly consistent with the observed trend (e.g., Harris 1991),
GCSs in some ellipticals show 
${\alpha}_{\rm gc} \approx {\alpha}_{\rm s}$
(e.g., Ashman \& Zepf 1999), which
is not fully consistent with the present results.

What additional physics is required to overcome these limitations?
As mentioned above, the inclusion of 
destruction processes may be important but this is unlikely to
explain the trends with host galaxy luminosity. 
The present study is dissipationless and so does not include the formation of the new GCs,
which would be formed preferentially in the central
regions of gaseous galaxy mergers 
along with {\it field} star formation (e.g. Bekki et al. 2002). 
A future study should investigate 
the role of new GC formation in shaping the radial density profiles
of GCSs. Given our lack of understanding of the physics of GC formation, this is
a difficult task. As we have noted above, the
contribution of new GCs to the existing GCSs of ellipticals
appears to be small.  

Bekki et al. (2003) showed that the strong tidal field
in galaxy clusters can strip GCs from cluster ellipticals 
and consequently steepen the radial density profiles of the
remaining GCS in the donor galaxy. Consequently, the recipient
galaxy may obtain a flatter GCS profile.

Recent theoretical and numerical studies suggest that
if cosmic reionization can strongly suppress the formation
of old, metal-poor GCs in dwarfs at high redshifts ($z>6$),
then the radial profiles of GCSs in galaxies would be influenced
at very early epochs (Santos 2003; Bekki 2005).



Finally, the accretion of GCs via minor mergers (discussed in the next section) may
also have an important effect on some galaxies. 

Thus the radial density
profiles of GCSs in galaxies could be influenced by 
several physical processes associated with galaxy formation
and evolution. 
Future studies should therefore seek to determine the 
{\it relative importance}  of each physical mechanism
in shaping the radial density profiles of GCSs. 
For this purpose, it is vital that future observational studies
provide observational data on
structural, kinematical, and chemical properties of GCs 
which can be compared with any theoretical predictions.
For example, systematic
observations of the radial dependence of GC luminosity functions
in ellipticals over a range of galaxy luminosities
will provide a clue as to the dominant physical processes.

   \begin{figure}
   \centering
   \includegraphics[width=8.5cm]{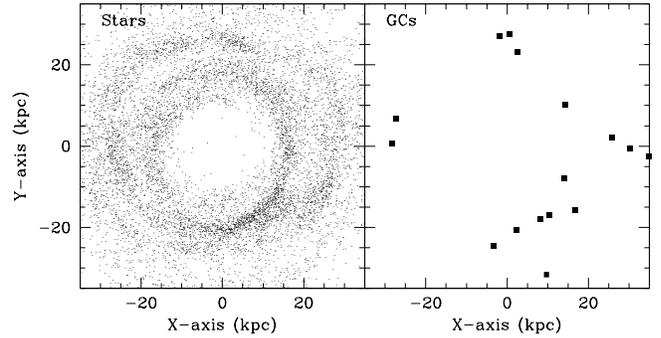}
      \caption{
Final distributions of stars (left)
and GCs (right) projected onto the $x$-$y$ plane for 
the disk galaxy model after minor merging of the disk with
a dwarf elliptical galaxy (dE) with $M_{\rm B}=-16$ mag. GCs in
the right panel for this minor merger model are those stripped from the dE
and represented by filled squares. 
The center of the disk galaxy with the initial size
of 17.5 kpc, which is not shown within 
the two frames for clarity, is coincident with the center 
of each frame.
Note that most GCs are located in the outer part of the disk.
              }
         \label{FigVibStab}
   \end{figure}

   \begin{figure}
   \centering
   \includegraphics[width=8.5cm]{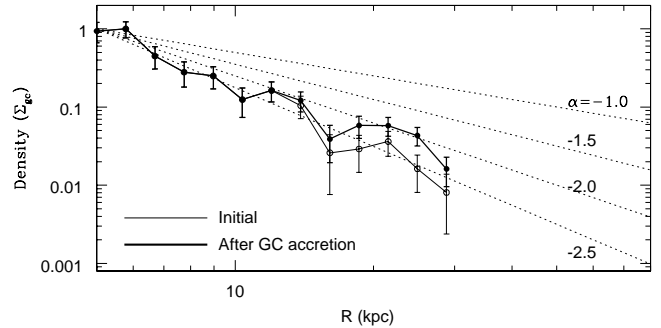}
      \caption{
The  projected number distributions 
and GCs in the initial disk model (left) 
and the minor merger one (right) where a dwarf elliptical
galaxy (dE) merges with the disk.
For clarity, the density distributions are normalized to
their central values.
Thin dotted lines represent power-law slopes ($\alpha$)
of $\alpha$ = $-2.5$, $-2.0$, $-1.5$, and $-1.0$.
Note that the outer profile ($R \sim 20$ kpc) of the GCS of the disk
becomes flattened after GC accretion associated with the minor merging.
              }
         \label{FigVibStab}
   \end{figure}

\subsection{Minor Mergers}

Although minor galaxy merging between a spiral and a dwarf 
can not form an elliptical 
(e.g., Walker et al. 1996; Bekki 1998), the accretion of GCs from the dwarf
onto the spiral (e.g. Forbes, Strader \& Brodie 2004) 
can possibly change the original density profiles of the GCS
of the spiral. 
Since full discussions on this matter are
obviously beyond the scope of this paper,
we here briefly discuss 
how GC accretion events 
can change the radial profiles of GCSs of galaxies
in a more general way.
Although the following discussions 
are on GC accretion onto disk galaxies,
{\it the basic  results}
can be also applied to GC accretion onto elliptical
galaxies (Bekki \& Forbes 2005).
Here we model a dwarf elliptical (dE) composed of dark matter, stars,
and GCs and investigate the dynamical evolution of GCs
during its orbital evolution around the present disk galaxy
model. The full details of the dE model orbiting
a spiral like the Galaxy or M31 can be found
in Bekki \& Freeman (2003),
Bekki \& Chiba (2004), and Bekki \& Forbes (2005).

Figure 7 shows the final distributions of stars and GCs initially
within a dE with $R_{\rm e}=0.8$ kpc,
$M_{\rm B}=-16$ mag,  and $S_{\rm N}=5$ (corresponding
to 20 GCs in the dE), where $S_{\rm N}$ is the specific frequency
of GCs.  In this model, the dE is assumed to be initially located
35 kpc from the center of the disk galaxy and have a circular
velocity of 209 km s$^{-1}$ at this position.  
The dE can be almost completely destroyed within $\sim 6$ Gyr
of its evolution around the disk galaxy, and GCs initially within
the dE are all tidally stripped and dispersed into the halo
region of the disk galaxy. The tidal destruction of the dE
can happen well outside the disk in this model so that
most GCs of the dE can be distributed in the outer part of
the disk's halo. Consequently, the radial density profile
of the GCSs composed both of the original GCs of the disk 
and of the stripped
GCs from the dE in the disk can become flattened, as shown
in Figure 8.

These results imply that GC accretion associated with
destruction of dwarfs in disk galaxies can flatten the
radial profiles of GCSs in the disk galaxies.
These results furthermore suggest that although
minor merging can not transform disk galaxies into
ellipticals (Walker et al. 1996; Bekki 1998), the radial profiles
of GCSs in disk galaxies can be significantly changed
by minor merger events if such events are repeated 
in the dynamical histories of the galaxies.
The density profiles of dark matter halos of galaxies
can be the key parameters in 
the flattening of the density profiles
of GCSs by GC accretion (Bekki \& Forbes 2005, in preparation).
Therefore the flattening processes 
by GC accretion are unlikely to depend on morphological
types (e.g., Sp, E, and cD) of luminous components,
if all galaxies have similar dark matter profiles like
the NFW profile.
Recent numerical simulations have suggested that GC accretion
from cluster member galaxies onto the central giant
cDs in clusters of galaxies is  highly likely 
(e.g., Bekki et al. 2003).
The results shown in figures 7 and 8
therefore can suggest that GC accretion can be also  responsible
for the origin of the very flat density profiles
of GCSs in cDs.

\section{Conclusions}

We have demonstrated that the observed
luminosity dependence of the GCS slope (${\alpha}_{\rm gc}$) 
can be reasonably well reproduced in elliptical galaxies
formed through  sequential dissipationless major merger events. 
Previous numerical simulations have already demonstrated that
the origin of structural non-homology in elliptical galaxies can be 
closely associated with the dynamics of major galaxy merging
(e.g., Capelato et al. 1995; Bekki 1998; Dantas et al. 2002).
We suggest that structural non-homology seen in {\it both} the stellar 
component and GCSs of ellipticals can  have a common origin,
i.e. the  growth of elliptical galaxies through
dissipationless major merging.
The derived $M_{\rm V}-r_{c}$
and ${\alpha}_{\rm gc}-{\alpha}_{\rm s}$ relationships 
are also the direct result of this  growth.
Thus the radial density profiles of GCSs in 
elliptical galaxies can be regarded
as containing fossil records of their  
merging histories. 
Given the fact that dynamical non-homology can be closely
associated with the origin of the Fundamental Plane 
(e.g., Djorgovski \& Davis 1987) of elliptical galaxies,
it is an interesting observational question whether
GCSs show an analogous ``Fundamental Plane''. 

\section*{Acknowledgments}
We are  grateful to the referee for valuable comments,
which contribute to improve the present paper.
K.B. and  DAF acknowledge financial support from the Australian
Research Council (ARC) throughout the course of this work. The
numerical simulations reported here were carried out on GRAPE
systems kindly made available by the Astronomical Data Analysis
Center (ADAC) at National Astronomical Observatory of Japan
(NAOJ).

{}

\end{document}